\documentclass[review, number]{elsarticle}
\graphicspath{ {./figures/} }
\usepackage{float}
\usepackage{verbatim} 
\usepackage{amsmath}
\usepackage{multirow}
\usepackage{array}
\usepackage{booktabs}
\restylefloat{figure}
\floatstyle{plaintop} 
\restylefloat{table}
\usepackage[margin=0.80in]{geometry}

\usepackage{ccaption}
\usepackage{amsmath}
\usepackage{amsbsy}
\usepackage{dcolumn}
\usepackage{multirow}
\usepackage[table]{xcolor}
\usepackage{wrapfig}
\usepackage{pifont}
\usepackage{float}
\usepackage{adjustbox}
\usepackage{wrapfig}
\usepackage{soul}
\usepackage{footnote}
\usepackage{times}
\usepackage{lipsum}
\usepackage{lineno}

\usepackage[T1]{fontenc}
\usepackage{textcomp}
\usepackage{graphicx}
\usepackage{graphics}
\usepackage{soul}

\usepackage{epstopdf}

\usepackage{caption}
\usepackage{subcaption}

\usepackage{amssymb,amstext,amsmath}
\usepackage{float}
\usepackage{booktabs}
\usepackage{array,ragged2e}
\usepackage[table]{xcolor}
\usepackage{multirow}
\usepackage{algorithm}
\usepackage{algpseudocode}
\usepackage{pifont}
\usepackage{enumerate}

\biboptions{square,numbers,sort&compress}
\journal{Digital Engineering}

\begin{document}
\begin{frontmatter}











\title{Analysis of Distracted Pedestrians Crossing Behavior: An Immersive Virtual Reality Application}

\author[label1]{Methusela Sulle \corref{cor1}}
\ead{msulle@scsu.edu}

\author[label1]{Judith Mwakalonge}
\ead{jmwakalo@scsu.edu}

\author[label2]{Gurcan Comert}
\ead{gcomert@ncat.edu}

\author[label1]{Saidi Siuhi}
\ead{ssiuhi@scsu.edu}

\author[label1]{Nana Kankam Gyimah}
\ead{ngyimah@scsu.edu}

\author[label1]{Jaylen Roberts}
\ead{jrober26@scsu.edu}

\author[label1]{Denis Ruganuza}
\ead{druganuz@scsu.edu}

\cortext[cor1]{Corresponding author.}
\address[label1]{Department of Engineering, South Carolina State University, Orangeburg, South Carolina, USA, 29117}
\address[label2]{Department of Computational Engineering and Data Science, North Carolina A\&T State University, Greensboro, North Carolina, US, 27411}

\begin{abstract}
Pedestrian safety is a critical public health priority, with pedestrian fatalities accounting for 18\% of all U.S. traffic deaths in 2022. The rising prevalence of distracted walking, exacerbated by mobile device use, poses significant risks at signalized intersections. This study utilized an immersive virtual reality (VR) environment to simulate real-world traffic scenarios and assess pedestrian behavior under three conditions: undistracted crossing, crossing while using a mobile device, and crossing with Light-emitting diode (LED) safety interventions. Analysis using ANOVA models identified speed and mobile-focused eye-tracking as significant predictors of crossing duration, revealing how distractions impair situational awareness and response times. While LED measures reduced delays, their limited effectiveness highlights the need for integrated strategies addressing both behavioral and physical factors. This study showcases VR’s potential to analyze complex pedestrian behaviors, offering actionable insights for urban planners and policymakers aiming to enhance pedestrian safety.
\end{abstract}


\end{frontmatter}

\section{Introduction}
\begin{figure*}[h!]
\centering
\includegraphics[scale=0.5]{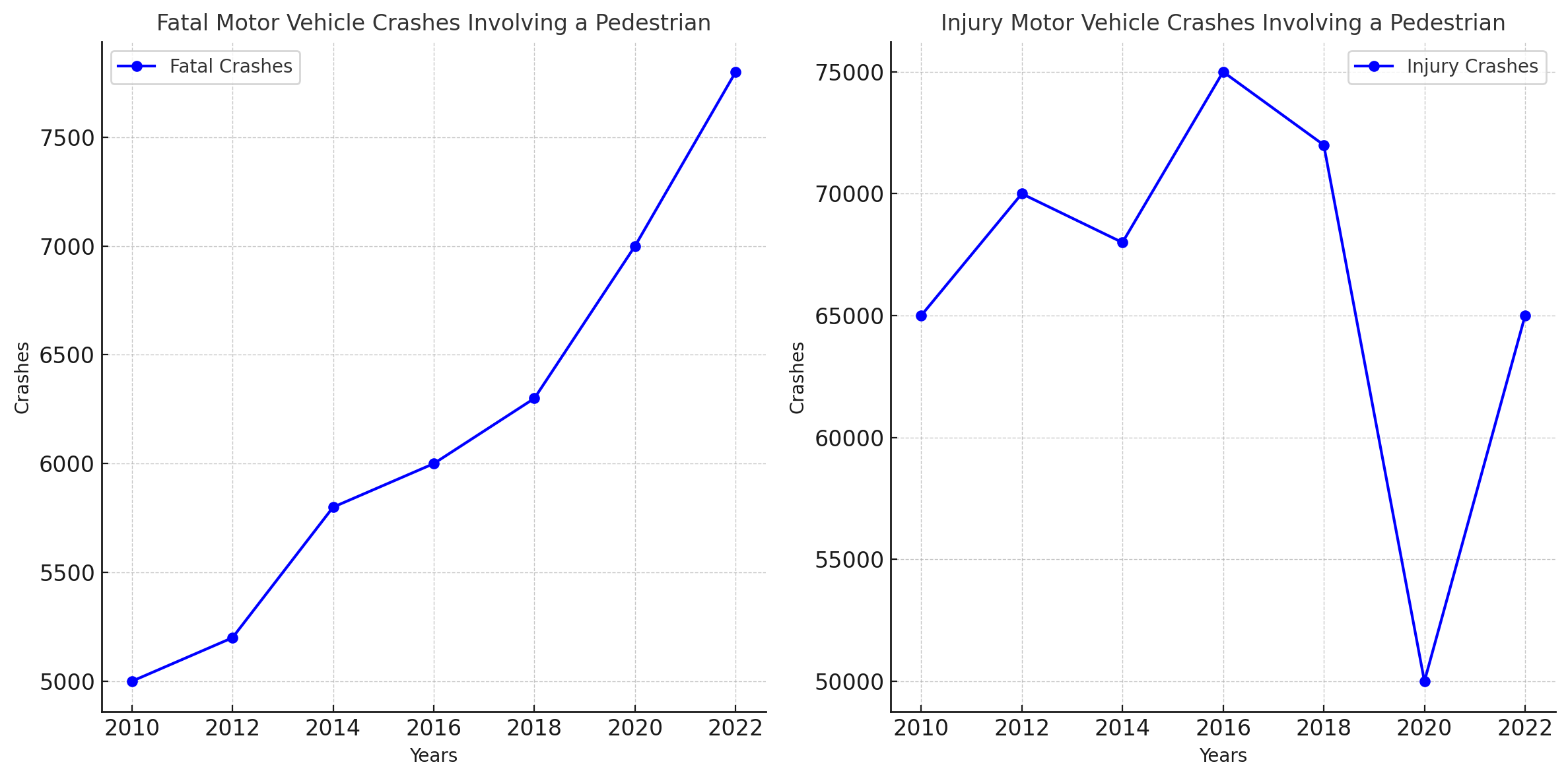}
\caption{Traffic Crashes Involving Pedestrians (2011–2021) \citep{nhtsaPedestrianSafety}: (a) Fatal crashes, (b) Injury crashes.}
\label{fig:traffic-statistics}
\end{figure*}

Pedestrian safety is a pressing public health issue, as nearly everyone is a pedestrian at some point \citep{retting2017pedestrian}. Despite the routine nature of walking, pedestrian injuries and fatalities remain a significant concern \citep{chakravarthy2007pedestrian, world2023pedestrian}. Over the past decade, pedestrian injuries have made up roughly 2.9\% of all traffic-related injuries in the U.S., while fatalities accounted for nearly 18\% of traffic deaths by 2022—the highest annual proportion recorded in over ten years \citep{ghsaPedestrianTraffic, radun2024systematic}. Figure \ref{fig:traffic-statistics}a and b shows a concerning rise in fatalities from 2011 to 2020, though injury rates have fluctuated. In 2020, 77\% of pedestrian fatalities occurred during dark conditions, with smaller percentages at dusk (2\%) and dawn (2\%). By 2021, urban areas had become particularly hazardous, with 84\% of fatalities occurring in these settings, 77\% on open roads, and 74\% in darkness, with or without artificial lighting \citep{nhtsaPedestrianSafety}. These trends underscore the urgent need for research addressing the complex and growing risks faced by pedestrians across diverse environments.

Compounding these risks is the widespread use of mobile devices, which has introduced significant challenges to pedestrian safety \citep{sobrinho2022risks}. Pedestrians are often distracted by activities such as social media, texting, phone calls, gaming, and navigation \citep{raoniar2024digital, chen2018smartphone, arafat2023effectiveness}. The portability of these devices enables such activities to occur virtually anywhere and at any time, yet this convenience comes at a cost: distractions reduce spatial awareness and heighten the likelihood of accidents \citep{stavrinos2018distracted, vankov2021effects}. Although some progress has been made in enhancing pedestrian safety, a significant gap remains in comprehensive data and standardized best practices for mitigating the specific risks posed by distracted walking \citep{mwakalonge2015distracted}. This gap highlights the necessity for targeted research on pedestrian behavior under distracted conditions, especially in high-risk areas like street crossings.

Distracted walking among pedestrians, particularly at street crossings, presents a critical safety issue \citep{o2023impact} that this research aims to address. Mobile device distractions impair pedestrian spatial awareness and decision-making, further increasing accident risks. However, analyzing distracted pedestrian behavior poses several significant challenges, which are outlined below: 
\begin{enumerate}
    \item Traditional methods for studying pedestrian behavior relies mostly upon the observations and road tests, which poses a significant safety risks to participants and lack the controlled environments needed for testing different scenarios \citep{schneider2020virtually}.
    
    \item Mobile-induced distractions impairs the spatial awareness of the pedestrians, leading to a slower crossing times, impact situational awareness, and larger the variability in their crossing speed, which can pose a significant risks to pedestrian safety \citep{boulagouas2024effects}.

    \item Existing safety measures, such as LED-illuminated crosswalks, only partially counteract the adverse effects of distracted pedestrians, highlighting the need for a more integrated approach to pedestrian safety \citep{hallewell2024field}.
\end{enumerate}

Previous studies have addressed these challenges with varying limitations. Kalatian and Farooq’s VR analysis of pedestrian behavior provided real-world insights but could not repeatedly simulate complex high-risk scenarios \citep{kalatian2018mobility}. Similarly, \citet{sobhani2018impact} model of smartphone distraction on pedestrians lacked the ability to fully simulate dynamic pedestrian-vehicle interactions. \citet{villena2022effects} demonstrated VR’s promise for studying distracted behaviors in controlled settings; however, VR’s application to pedestrian safety research remains limited, underscoring the need for a more versatile framework.

To address the challenges above, this study proposed a virtual immersive reality environment to analyze pedestrian behaviors in several scenarios, including Undistracted crossing, mobile phone distraction, and LED safety interventions at marked crosswalks at the signalized intersections. This framework models the real traffic conditions, which offers a controlled, safe, and repeatable scenarios for data collection. It incorporates state-of-the-art tracking technologies like eye-tracking, enabling capture of precise behavioral data and providing insights into pedestrian responses while minimizing risks associated with field studies. The main contributions of this framework directly address the challenges outlined above: 
\begin{enumerate}
    \item This study introduces a VR-based framework that models real-world intersections with realistic traffic patterns and incorporates scenarios with varying distraction levels and safety measures, demonstrating its effectiveness as a safe and controlled alternative for studying pedestrian behavior.

    \item We quantify the effects of mobile-induced distractions on pedestrian behavior, revealing impaired spatial awareness, slower crossing times, reduced situational awareness, and increased variability in crossing speed. These findings underscore the significant safety risks posed to pedestrians.

    \item We adopted an integrated approach to pedestrian safety by combining LED innovations with behavioral interventions, effectively addressing the limitations of existing measures, such as LED-illuminated crosswalks, which only partially mitigate the risks posed by distracted pedestrians.
\end{enumerate}

The rest of this paper is organized as follows: Section $\text{II}$ provides a summary of recent relevant studies that applied virtual reality techniques to analyze pedestrians’ distracted walking at intersections. Section $\text{III}$ describes the methodology which outlines the research design, data collection, and data analysis processes used to examine pedestrian behavior under distraction. Section $\text{IV}$ presents the experimental results and discussion. Finally, the conclusion and future work are provided in Section $\text{V}$.

\section{Related Works}
Research on pedestrian-distracted walking safety typically focuses on several themes, including types of distractions, simulation methods, safety measures, and counteractive interventions. A summary of the literature along these lines is presented below, along with a critical evaluation of each study’s contributions and limitations, highlighting gaps this study seeks to address.

\subsection{Distraction Types and Behavior Analysis}
Pedestrian behavior can vary based on several factors, including Age, Gender, Group size, and other environmental factors. In the study, \citet{https://publications.polymtl.ca/9169/} examined the pedestrian safety behavior at urban signalized intersections using Logistic regression the author highlighted factors influencing risky behaviors like crossing on red lights, emphasizing the role of intersection design, pedestrian characteristics, and wait times in safety outcomes. Using the observational data collection method from 24 intersections (12 per city) in Montreal and Quebec City the study found that the most common distractions include using headphones and walking with companions, while significant head movement differences were observed between cities. 
\citet{russo2018pedestrian} employed Ordinary Least Squares (OLS) regression to model pedestrian walking speeds and binary logit models to analyze pedestrian distraction and violations to investigate pedestrian behaviors at signalized crosswalks, focusing on distracted walking, pedestrian signal violations, and walking speeds. Utilizing field-recorded observational data from one intersection in New York and three in Flagstaff, Arizona, the researchers analyzed data from 3,038 pedestrian crossings through video-based observation and found that specific site and demographic variables significantly influenced pedestrian distraction, signal violations, and walking speeds. On the other hand, \citet{thompson2013impact} investigated the influence of social and technological distractions on pedestrian crossing behavior, focusing on cautionary actions and crossing durations. Using a prospective observational approach, data were collected at 20 high-risk intersections in Seattle, recording demographic, behavioral, and distraction data of 1102 pedestrians. Key data collected were distraction types such as texting, talking, or listening to music. Results showed that nearly 30\% of pedestrians engaged in distractions while crossing, with text messaging notably increasing crossing times and the likelihood of risky behaviors, like failing to obey traffic signals.
 Furthermore, \citet{kalatian2018mobility} observed pedestrian behaviors in real environments, focusing on socio-demographic influences like age and gender on waiting times before crossing. While this study provided real-world insights, its observational nature could not repeatedly simulate high-risk scenarios for detailed analysis. \citet{schwebel2012distraction} examined college students crossing a virtual street while engaging in distracting activities (e.g., texting, talking, listening to music) and found that distracted participants were more likely to be "hit" in the virtual simulation. 

\subsection{Methodological Approaches: Simulation Environments}
A common approach for studying distracted walking is using simulated environments to observe pedestrian responses under various scenarios.  \citet{stavrinos2009effect, stavrinos2011distracted} used immersive virtual environments to assess children and college students crossing streets while on cell phones, highlighting significant declines in attention and riskier behaviors. However, while immersive, these studies did not examine multiple types of distractions or compare different safety measures within the same simulation, limiting their applicability to broader distracted pedestrian contexts. Virtual Immersive Reality Environments (VIREs) are emerging as tools for testing diverse distraction scenarios, allowing researchers to simulate complex, controlled environments safely \citep{villena2022effects}. 
\citet{wang2022effect} employed a descriptive statistical and ANOVA analysis methods in assessing crossing behaviors under varying traffic conditions (easy vs. hard) for participants characterized by high or low sensation-seeking tendencies. The study, used virtual reality (VR) environment, focusing on children (10-13 years), adolescents (14-18 years), and young adults (20-24 years), Data collection included 209 participants who completed virtual crossings under both traffic scenarios. Key findings revealed that adolescents, particularly those high in sensation-seeking, exhibited the most dangerous crossings. Age and sensation-seeking were significant predictors of risky pedestrian behavior, with traffic conditions significantly influencing crossing behaviors across groups.
Furthermore, \citet{deb2017efficacy} evaluated the efficacy of using virtual reality (VR) technology for pedestrian safety research by developing a pedestrian simulator using the HTC VIVE headset and Unity software. The study involved, tracking participants' head movements as they navigated a virtual intersection with varying traffic scenarios, measuring their responses, and gathering subjective data on user experience and simulation sickness. Data from 21 participants, excluding four due to simulator sickness, were analyzed for crossing behaviors and collision rates under different traffic conditions. Key findings showed that VR environments replicated realistic pedestrian behaviors, with participants adjusting their speed and actions based on traffic conditions, confirming VR's potential as a valid tool for studying pedestrian safety behaviors in controlled yet realistic settings.

\subsection{Safety Performance Metrics and Surrogate Measures}
Many studies evaluate pedestrian safety through measures like Time to Collision (TTC), Post-Encroachment Time (PET), and maximum acceleration or deceleration. \citet{sobhani2018impact} explored the effects of smartphone use on pedestrian wait times and decision-making, finding that distracted individuals tend to delay crossing, potentially affecting traffic flow. \citet{gaarder1989pedestrian} and \citet{ni2016evaluation} utilized conflict indicators like Time to Collision (TTC) and Post-Encroachment Time (PET) to assess pedestrian safety at signalized intersections, using observational data collection method, Support Vector Machine (SVM) was used in the analysis of the data, and the key findings were that signalization generally reduces pedestrian risk, with greater efficacy at high-speed intersections and with limited turning traffic, while factors such as crosswalk proximity, refuge presence, and red-light behavior influenced safety outcomes and that conflict indicators are good for better accuracy in severity estimation, suggesting improvements in pedestrian safety analysis and targeted traffic behavior education.
However, these studies approach lacked the ability to simulate real-time pedestrian-vehicle interactions, limiting its relevance for high-stakes scenarios. The research suggests a need for more robust experimental designs to validate these measures in diverse conditions.

\subsection{Counteractive Measures and Intervention Effectiveness}
Countermeasures such as LED flashlights and crossing reminders have been proposed to mitigate distracted walking risks. One study \citet{kalatian2018mobility} tested LED crosswalk lights as a safety measure, noting a slight increase in crossing success rates but no significant reduction in overall risk, as the lights may have contributed to distraction. \citet{basch2014technology} studied distracted walking behaviors at high-risk intersections in Manhattan, observing that nearly 30\% of pedestrians were distracted by mobile devices regardless of traffic signals. While the findings highlighted the prevalence of distraction, the study’s observational nature limited the exploration of practical interventions to reduce it.

The aforementioned literature indicates that distractions significantly reduce pedestrian safety, yet effective and scalable interventions remain limited. Few studies have examined countermeasures in controlled virtual environments. To address these gaps, this study employs a Virtual Immersive Reality to simulate pedestrian behavior across three distinct scenarios: Undistracted crossing, crossing while using a handheld device, and crossing with an implemented safety measure. These simulations provide a versatile framework for developing and evaluating interventions, offering practical insights for real-world applications.

\section{Methodology}
This section outlines the research design, data collection, and data analysis processes used to examine pedestrian behavior under distraction. Each methodological step is aligned with the study’s primary objective: to assess the impact of distractions on pedestrian safety and explore practical interventions using a Virtual Immersive Reality Environment.

\subsection{Study Design and Simulated Environment}
\begin{figure}[H]
\centering
\begin{subfigure}{0.499\columnwidth}
\centering
  \includegraphics[width=0.99\textwidth]{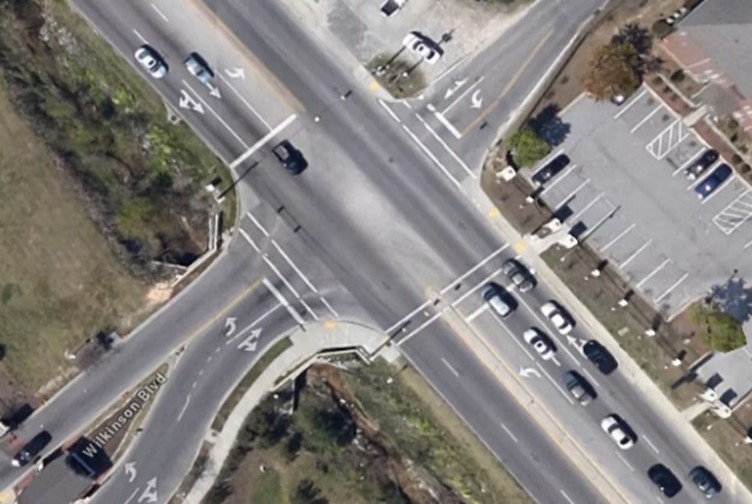}
\caption{Satellite image of the real-world traffic intersection at coordinates 33.50260, -80.84521, used to model the simulated environment.}
  \label{fig:satelite}
\end{subfigure}%
\begin{subfigure}{0.499\columnwidth}
 \centering
\includegraphics[width=0.99\textwidth]{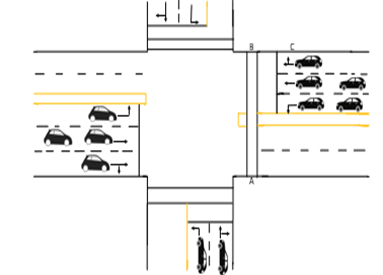}
  \caption{A figure of the intersection, showing Point A as the starting point, Point B as the ending point, and Point C as the location of the billboard for the student village.}
  \label{fig:sketched-intersect}
\end{subfigure}
\caption{Intersection Studied.}
\end{figure}

The pedestrian crossing model in the Virtual Reality (VR) environment was developed by integrating a 3D simulation created with Unity, Blender, and scripts written in VS Code. Unity is a widely used 3-D game engine for developing video games, animations, and other 3-D applications, was employed to design immersive virtual scenarios that simulate pedestrian crossings with realistic traffic patterns, behavioral dynamics, and distraction conditions \citep{brookes2020studying}. Blender, a program to project and render 3-dimensional objects,  was utilized to create 3D assets, such as buildings, vehicles, and crosswalk elements, which significantly enhanced the environmental realism of the simulation. VS Code was used to script interactive elements, coordinate real-time data collection, and synchronize behaviors within the VR simulation. Firebase was implemented for data persistence and management, ensuring the reliable tracking and analysis of key participant behavior metrics across various simulation scenarios. The modeled environment replicates the intersection at US-456 near South Carolina State University, selected for its relevance in studying distracted walking behaviors among young adults who frequently cross this location. Key features, including traffic flow, pedestrian signals, and bidirectional vehicle movement, were simulated to accurately reflect real-world conditions. Fig. \ref{fig:satelite} shows a satellite image of the location, while Fig. \ref{fig:sketched-intersect} presents a sketch of the intersection, detailing pedestrian entry and exit points as well as vehicle pathways.

\begin{figure}[H]
\centering
\begin{subfigure}{0.45\columnwidth}
\centering
  \includegraphics[width=0.99\textwidth]{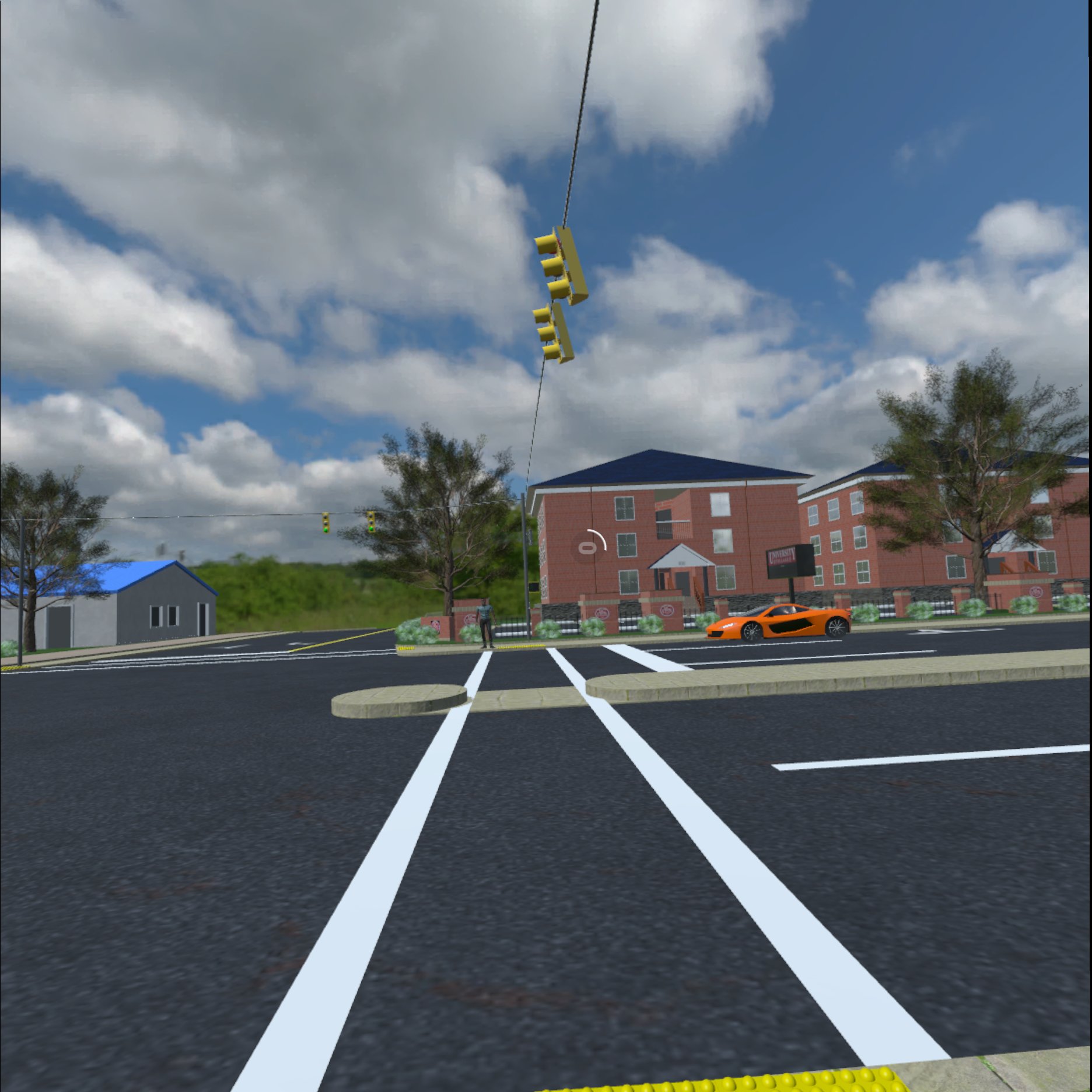}
\caption{Simulated crossing to collect data for the first two scenarios.}
  \label{fig:simulated-2nd}
\end{subfigure}%
\begin{subfigure}{0.45\columnwidth}
 \centering
\includegraphics[width=0.99\textwidth]{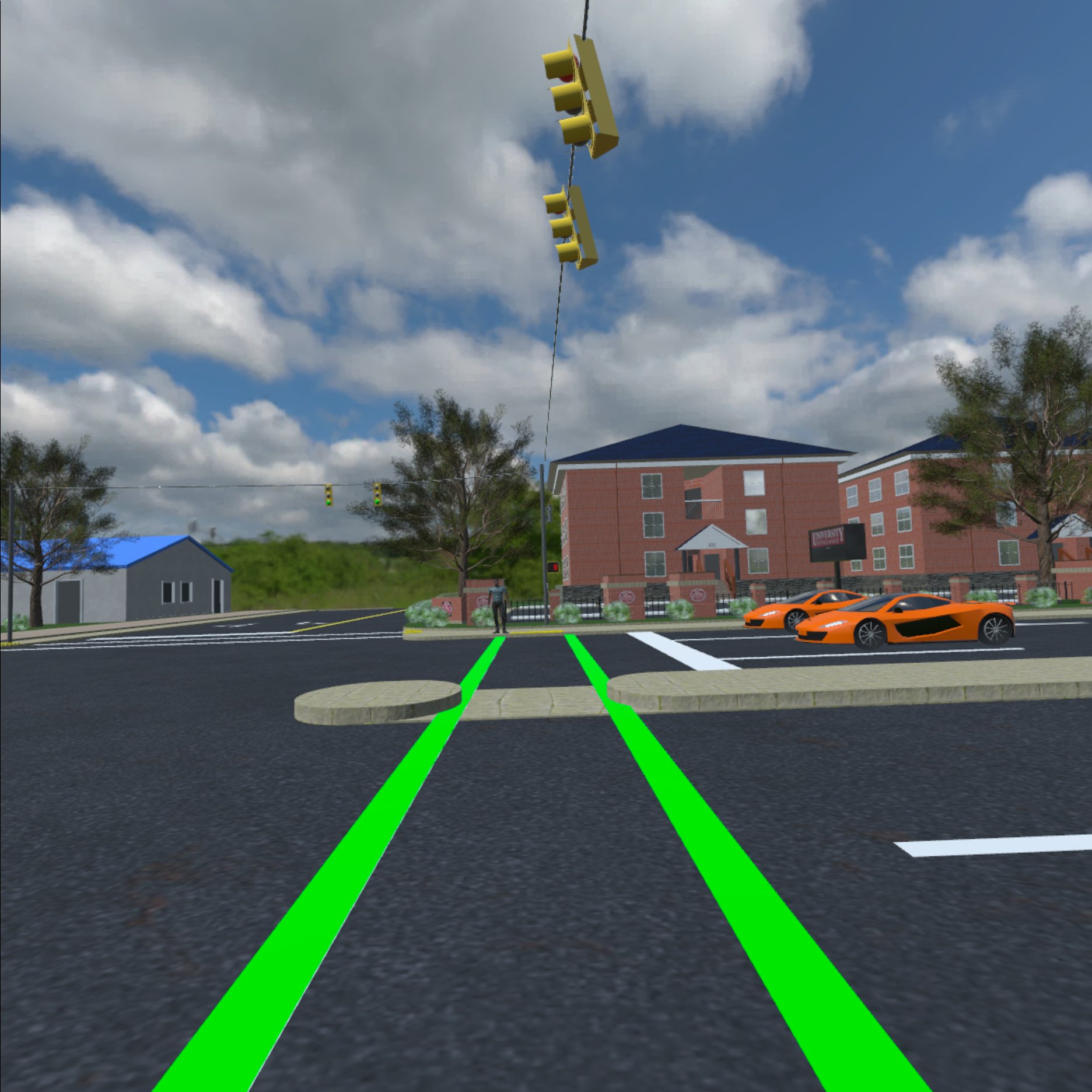}
  \caption{Simulated crossing to collect data for the for the third scenario.}
  \label{fig:simulated-third}
\end{subfigure}
\caption{Simulated Scenarios.}
\end{figure}

Within this realistic setting, the study analyses the outcomes of pedestrian crossing behaviors with and without presence of the distraction situations using VR equipment. Fig. \ref{fig:simulated-2nd} and Fig. \ref{fig:simulated-third} are the virtual environments used in our study. Fig. \ref{fig:simulated-2nd} will collect data for the first two scenarios, while Fig. \ref{fig:simulated-third} will be utilized for the third scenario. Three distinct scenarios were developed to test different levels of pedestrian distraction:
\begin{enumerate}
    \item \textbf{Undistracted Crossing}: A baseline scenario in which participants cross without any distractions.
    \item \textbf{Crossing While Using a Handheld Device}: Participants cross while using a simulated smartphone, representing typical pedestrian distractions.
    \item \textbf{Crossing with Safety Measures (LED Light)}: Participants cross while distracted by a smartphone, with an LED safety light which detects and activates as a pedestrian stepped on the crosswalk to alert them to traffic, thereby testing the effectiveness of an intervention.
\end{enumerate}

These scenarios were designed to assess how varying distraction levels impact pedestrian decision-making and safety, directly supporting the study’s objective of evaluating intervention strategies for distracted walking.

\hspace{1cm}
\subsection{Data Collection and Study Framework}
To comprehensively assess pedestrian behavior under distraction, this study utilized a realistic virtual environment modeled after a real-world intersection, along with immersive VR equipment, to capture detailed behavioral data across varied distraction scenarios.

\subsubsection{Immersive VR Application}
\begin{figure}[H]
\centering
\begin{subfigure}{0.6\columnwidth}
\centering
  \includegraphics[width=0.99\textwidth]{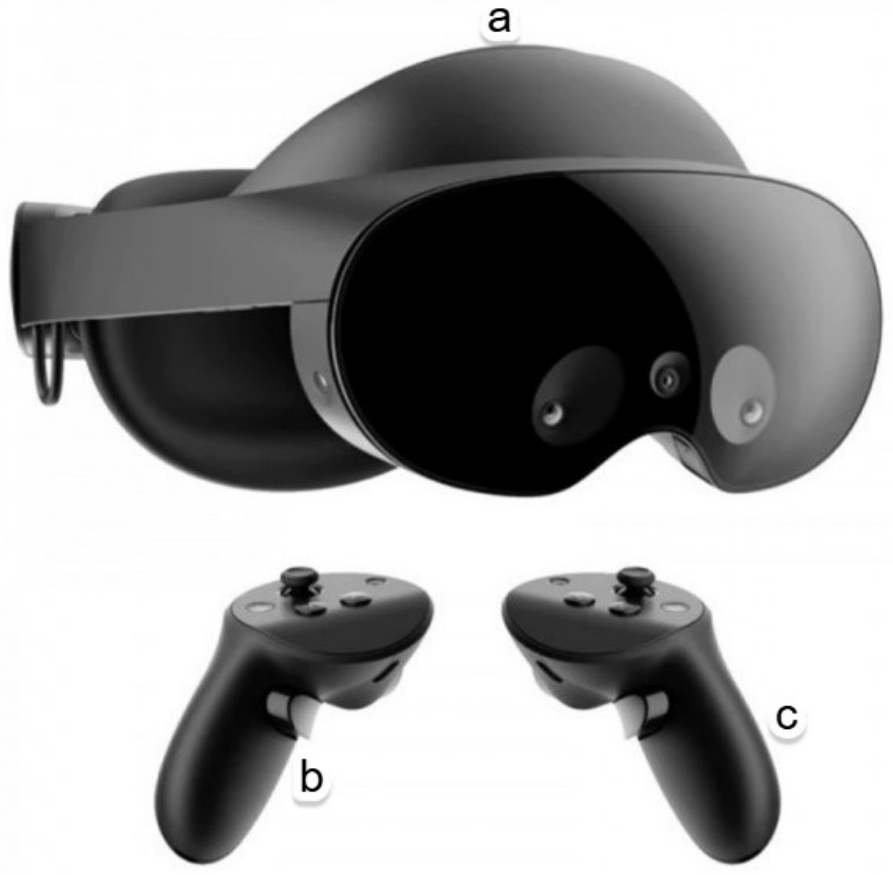}
\caption{VR headset used for data collection.}
  \label{fig:vr-headset}
\end{subfigure}%
\begin{subfigure}{0.45\columnwidth}
 \centering
\includegraphics[width=0.7\textwidth]{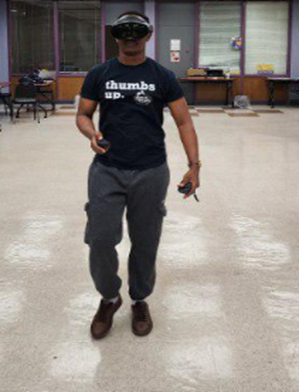}
  \caption{Participant in the study wearing virtual reality headsets as they engage in simulated crossing scenarios.}
  \label{fig:vr-part1}
\end{subfigure}
\caption{Data Collection.}
\end{figure}

The immersive VR simulation was conducted using a Meta Quest Pro headset, selected for its high-resolution display and precise tracking capabilities. This equipment enabled accurate data collection on participants' visual attention, crossing speed, and responses to the LED light intervention. The headset’s built-in eye-tracking sensors recorded distraction metrics, such as gaze direction and frequency, providing essential data on pedestrian distraction levels. Fig. \ref{fig:vr-headset} illustrates the VR headset used in the study, where it consists of three components: (a) the headset itself, which serves as the primary device for immersive visual and auditory experiences; (b) the right controller, utilized by pedestrians for writing in the virtual smartphone environment; and (c) the left controller, which functions to bring up and toggle the virtual smartphone on and off \citep{metaquestpro}. 

\subsubsection{Data Collection}
\begin{enumerate}
\item \textbf{Pre-Experiment}: The data collection process involved gathering both field and VR data, with 6 participants recruited from South Carolina State University and additional participants recruited in collaboration with Benedict College, Columbia, SC, to ensure diversity in age, gender, and phone usage habits. Each participant completed 8 runs, which were averaged to generate representative values for comparing VR and field data. This preliminary step aimed to ensure that each dataset reflects typical crossing durations under both real-world and virtual environments, laying a solid foundation for statistical analysis. 

\item \textbf{Actual-Experiment}: In total, $29$ participants, aged $15-25$ years, were selected to represent a demographic typical of young adult pedestrians; a group prone to distracted behaviors and at elevated risk for pedestrian accidents. Selection criteria included prior mobile device experience to ensure familiarity with smartphone interactions. Table \ref{Demographic Information} summarizes participant demographics categorized by gender and age group.
\end{enumerate}

\begin{table}[h!]
\centering
\caption{Participant demographics categorized by gender and age group, with counts and percentages.}
\scalebox{0.8}{
\begin{tabular}{|c|c|c|}
\hline
\textbf{Category} & \textbf{Subcategory} & \textbf{Count} \\
\hline
\multirow{2}{*}{Gender} & Male & 15 \\
 & Female & 14 \\
\hline
\multirow{3}{*}{Age} & 15-19 & 7 \\
 & 20-24 & 20 \\
 & 25 - Above & 2 \\
\hline
\end{tabular}
}
\label{Demographic Information}
\end{table}

\subsubsection{Operational Variables}
Data collection focused on capturing key behavioral metrics related to pedestrian distraction across different scenarios. For each scenario, participants’ wait times, crossing times, and crossing speeds were recorded in real time using the VR application, along with eye-tracking metrics to assess visual engagement, as summarized in Table \ref{Variables Data Collected}. Data were securely stored and analyzed to maintain participant confidentiality. Fig. \ref{fig:vr-part1} illustrates the VR setup used during data collection.

The primary variables measured in this study are defined as follows:
\begin{itemize}
    \item \textbf{Wait Time}: The duration from arrival at the crossing until the participant begins to cross.
    \item \textbf{Crossing Time}: The time taken to cross from one side of the road to the other.
    \item \textbf{Crossing Speed}: The average speed at which participants cross the road.
    \item \textbf{Eye Tracking}: The participant’s gaze direction and duration, indicating visual engagement and potential distraction.
\end{itemize}

\begin{table}[h!]
\centering
\caption{Summary of variables collected during the study.}
\renewcommand{\arraystretch}{1}
\scalebox{0.9}{
\begin{tabular}{| l |}
\hline
Gender \\
\hline
Age \\
\hline
EyeTracking\_Building (s) \\
\hline
EyeTracking\_Car (s) \\
\hline
EyeTracking\_Mobile\_Crossing (s) \\
\hline
EyeTracking\_People (s) \\
\hline
EyeTracking\_Road\_Crossing (s) \\
\hline
Crossing\_duration (s) \\
\hline
Crossing\_speed (fps)\\
\hline
Final\_walk\_speed (fps)\\
\hline
Initial\_walk\_speed (fps)\\
\hline
Number\_of\_glance\_during\_crossing \\
\hline
Using\_mobile\_duration\_at\_waitpoint (s)\\
\hline
Wait\_time\_duration (s) \\
\hline

\end{tabular}
}
\label{Variables Data Collected}
\end{table}

These metrics are essential for evaluating the effects of distractions and interventions on pedestrian behavior, providing a comprehensive assessment of how participants interact with the simulated environment.

\subsubsection{Pre and Post-Study Survey}
To complement the behavioral data, participants completed both pre- and post-study surveys to capture demographic information, phone usage habits, and subjective feedback on their VR experience. The pre-survey data in Table \ref{pre-survey part A} provides insights into participants' ages, phone usage, and distracted walking behaviors, aligning with the study’s structured approach to evaluating distraction impacts and potential interventions for pedestrian safety.

\begin{table*}[h!]
    \centering
    \caption{Pre-survey data on participants’ demographics, phone usage habits, and behaviors related to walking distractions and crosswalk safety.}
    \footnotesize
    \renewcommand{\arraystretch}{1} 
    \setlength{\tabcolsep}{5pt} 
    \scalebox{0.8}{
    \begin{tabular}{|>{\raggedright\arraybackslash}p{4.5cm}|>{\raggedright\arraybackslash}p{4.5cm}|c|c|}
        \hline
        \textbf{Questions} & \textbf{Categories} & \textbf{Frequency} & \textbf{Percentage} \\ \hline
      
        Age & 15-19 & 7 & 24\% \\ \cline{2-4}
            & 20-24 & 20 & 69\% \\ \cline{2-4}
            & 25 - Above & 2 & 7\% \\ \hline
        Gender & Male & 15 & 52\% \\ \cline{2-4}
               & Female & 14 & 48\% \\ \hline
        \multirow{5}{4.5cm}{How many years have you owned a cell phone?} 
            & Less than a year & 0 & 0\% \\ \cline{2-4}
            & 1-2 years & 1 & 3\% \\ \cline{2-4}
            & 3-5 years & 4 & 14\% \\ \cline{2-4}
            & 6-9 years & 11 & 38\% \\ \cline{2-4}
            & More than 9 years & 13 & 45\% \\ \hline
        \multirow{4}{4.5cm}{How many hours a day are you on your phone?} 
            & Less than one hour a day & 0 & 0\% \\ \cline{2-4}
            & 1-2 hours a day & 2 & 7\% \\ \cline{2-4}
            & 3-5 hours a day & 9 & 31\% \\ \cline{2-4}
            & More than 5 hours a day & 18 & 62\% \\ \hline
        \multirow{6}{4.5cm}{What takes up most of your total phone usage?} 
            & Social media & 15 & 52\% \\ \cline{2-4}
            & Email usage & 3 & 10\% \\ \cline{2-4}
            & Texting & 2 & 7\% \\ \cline{2-4}
            & Making phone calls & 2 & 7\% \\ \cline{2-4}
            & Phone gaming & 0 & 0\% \\ \cline{2-4}
            & Music listening & 7 & 24\% \\ \hline
        \multirow{5}{4.5cm}{How often do you use your phone while walking?} 
            & Frequently & 5 & 17\% \\ \cline{2-4}
            & Usually & 10 & 34\% \\ \cline{2-4}
            & Sometimes & 9 & 31\% \\ \cline{2-4}
            & Rarely & 5 & 17\% \\ \cline{2-4}
            & Never & 0 & 0\% \\ \hline
        \multirow{4}{4.5cm}{How often have you seen people walking distracted by cell phones on campus this week?} 
            & Very often & 9 & 31\% \\ \cline{2-4}
            & Often & 15 & 52\% \\ \cline{2-4}
            & Not often & 1 & 3\% \\ \cline{2-4}
            & Never & 4 & 14\% \\ \hline
        \multirow{4}{4.5cm} {I typically use a crosswalk}
            & 0 days a week & 2 & 7\% \\ \cline{2-4}
            & 1-2 days a week & 12 & 41\% \\ \cline{2-4}
            & 3-5 days a week & 8 & 28\% \\ \cline{2-4}
            & 6-7 days a week & 7 & 24\% \\ \hline
        \multirow{4}{4.5cm} {How cautious are you while using crosswalks?} 
            & Very cautious & 18 & 62\% \\ \cline{2-4}
            & Cautious & 8 & 28\% \\ \cline{2-4}
            & Somewhat cautious & 3 & 10\% \\ \cline{2-4}
            & I could be more cautious. & 0 & 0\% \\ \hline
        \multirow{2}{4.5cm} {Have you had a near-miss or accident while distracted as a pedestrian?}
            & Yes & 12 & 41\% \\ \cline{2-4}
            & No & 17 & 59\% \\ \hline
        \multirow{4}{4.5cm} {Why do people text and walk or drive distracted despite knowing the safety risks?}
            & \multicolumn{3}{l|}{For Personal gratification }\\ \cline{2-4}
            & \multicolumn{3}{l|}{In emergencies/urgent situations}\\ \cline{2-4}
            & \multicolumn{3}{l|}{Overconfidence}\\ \cline{2-4}
            & \multicolumn{3}{l|}{Negligence of risks}\\ \cline{2-4}
            & \multicolumn{3}{l|}{As a New norm}\\ \cline{2-4}
            & \multicolumn{3}{l|}{Addiction to smartphones}\\ \cline{2-4}
            & \multicolumn{3}{l|}{No laws restricts pedestrian phone use}\\ \hline
        \multirow{4}{4.5cm} {How familiar are you with virtual reality?}
            & Very familiar & 4 & 14\% \\ \cline{2-4}
            & Familiar & 7 & 24\% \\ \cline{2-4}
            & Somewhat familiar & 12 & 41\% \\ \cline{2-4}
            & Not familiar & 6 & 21\% \\ \hline
    \end{tabular}
    }
    \label{pre-survey part A}
\end{table*}

Post-study feedback, shown in Table \ref{Post-survey}, reveals that smartphone use was the most challenging distraction for participants to navigate. Participants also suggested several enhancements to improve realism in future VR simulations, such as adding group crossings and background elements (e.g., bicyclists, pets, or children), and incorporating additional distractions like listening to music. Many participants described the VR experience as realistic and emphasized its potential to improve pedestrian safety by providing insights into distraction-related behaviors and supporting interventions, such as automatic red lights to stop vehicles when pedestrians unexpectedly cross.





%

\begin{table}[h!]
\centering
\caption{Summary of post-survey participant feedback on distractions, VR improvements, experience impressions, and real-world applications.}
\renewcommand{\arraystretch}{1} 
\scalebox{0.8}{
\begin{tabular}{|>{\raggedright\arraybackslash}p{6cm}|>{\raggedright\arraybackslash}p{8cm}|}
\hline
\textbf{Questions} & \textbf{Responses} \\ \hline

\multirow{2}{6cm}{What distractions were most challenging while crossing the street?} 
& Smartphone usage \\ \cline{2-2}
& LED lights grabbing attention \\ \hline

\multirow{2}{7cm}{What improvements could enhance the VR simulation's realism and effectiveness?} 
& Adding bicyclists in the VR \\ \cline{2-2}
& A group of people crossing scenario \\ \hline

\multirow{3}{6cm}{What distractions or scenarios should future VR iterations include?} 
& Listening to music while crossing \\ \cline{2-2}
& People with animals (pets) \\ \cline{2-2}
& Crossing with children \\ \hline

\multirow{2}{6cm} {What are your impressions of the VR pedestrian crossing simulation?}
& Felt real \\ \cline{2-2}
& Great experience \\ \hline

\multirow{3}{6cm} {How can VR insights improve real-world pedestrian safety?}
& No physical harm to participants \\ \cline{2-2}
& Helps understand pedestrian distraction \\ \cline{2-2}
& Automatic vehicle stop lights for unexpected pedestrian crossings \\ \hline

\end{tabular}
}
\label{Post-survey}
\end{table}

These survey responses enhance our understanding of the VR data, providing context on real-world phone usage and participants' perceptions of distraction. This feedback underscores the importance of VR-based research in pedestrian safety, with participants recognizing VR’s potential to simulate realistic scenarios and suggesting improvements to make future studies even more applicable to real-world conditions.

\section{Results and Discussion}
This section presents key findings on pedestrian behavior under distraction, focusing on the impact of smartphone use on crossing durations, walking speed variability, and the effectiveness of LED safety interventions. Results from the VR simulations and field comparisons offer insights into the effects of distraction on pedestrian safety, emphasizing both the potential and limitations of safety measures like LED lights.

\subsection{Validation and Reliability of VR generated data}

\subsubsection{VR Environment Validation for Pedestrian Behavior Analysis}
To establish the VR environment’s reliability as a tool for analyzing distracted pedestrian behavior, we conducted a statistical test to compare VR data with field data and determine if the two methods of data collection significantly differ. The test focused on evaluating pedestrian crossing duration. Six participants completed 8 crossing trials in both environments. Results showed an average crossing duration difference of 0.31 seconds, with a standard deviation of 1.81 and a t-statistic of 0.41 (Table \ref{tab:data_comparison}). These values are below the critical t-value of 2.571 for 5 degrees of freedom, indicating no statistical significant difference between VR and real-world performances. This finding supports the VR model's validity for studying pedestrian behavior under distraction scenarios.

\begin{table}[h!]
    \centering
    \caption{Summary of statistical t-test results comparing field data and VR data for pedestrian crossing behavior, including average differences, standard error, and t-statistics.}
    \scalebox{0.7}{
    \begin{tabular}{>{\centering\arraybackslash}p{1.5cm}>{\centering\arraybackslash}p{1.5cm}>{\centering\arraybackslash}p{1.5cm}>{\centering\arraybackslash}p{1.5cm}}
        \textbf{User Pair} & \textbf{Field Data} & \textbf{VR Data} & \textbf{Difference (d)} \\
        \midrule
        1 & 19.75 & 16.88 & -2.87 \\
        2 & 18.47 & 21.13 & 2.65 \\
        3 & 17.94 & 19.00 & 1.06 \\
        4 & 17.97 & 18.13 & 0.16 \\
        5 & 16.96 & 17.13 & 0.16 \\
        6 & 17.82 & 18.50 & 0.68 \\
        \midrule
        \multicolumn{2}{l}{\textbf{Average of d:} 0.31} &
        \multicolumn{2}{l}{\textbf{Standard deviation of d:} 1.81} \\
        \multicolumn{2}{l}{\textbf{Standard error of d (SE):} 0.74} &
        \multicolumn{2}{l}{\textbf{t statistic:} 0.41} \\
        \multicolumn{2}{l}{\textbf{$t_{5, 0.025}$ (p-Value):} 2.571} & & \\
        \bottomrule
    \end{tabular}
    }
    \label{tab:data_comparison}
\end{table}

\subsubsection{Model Validation and Variable Significance}
To assess the VR model’s ability to predict pedestrian behavior under different distraction scenarios, we employed one way ANOVA statistical method to analyze all three conditions using F-statistics and p-values for each variable (Table \ref{ANOVA Model}). Key variables, including Final Walk Speed, Eye-Tracking Mobile Crossing, and Crossing Speed, showed strong predictive value with p-values of 0.002, 0.004, and 0.010, respectively. These findings confirm that these variables capture critical aspects of distracted behavior.

In contrast, variables such as Initial Walk Speed and Wait Time Duration had higher p-values, indicating weaker statistical support, although they still contribute to the overall model’s framework. Additional variables—Age, Gender, and peripheral eye-tracking metrics (e.g., buildings, cars)—showed less individual influence but maintained overall model robustness, as confirmed by normality checks across variables.

\begin{table*}[h!]
\centering
\caption{ANOVA model validation results with F-statistics, p-values, residual sums, and normality checks for crossing behavior variables.}
\renewcommand{\arraystretch}{1} 
\scalebox{0.8}{
\begin{tabular}{|>{\raggedright\arraybackslash}p{6.5cm}|c|c|c|c|}
\hline
\textbf{Variable}& \textbf{Sum Sq (Q)}& \textbf{F-Statistic (Q)}& \textbf{Residual Sum Sq}& \textbf{p-value}\\ \hline
Final\_Walk\_Speed (fps) & 58.577& 10.324& 482.274& 0.002\\ \hline
EyeTracking\_Mobile\_Crossing (s)
& 51.453& 8.937& 489.398& 0.004\\ \hline
Crossing\_Speed (fps)
& 41.339& 7.035& 499.514& 0.009\\ \hline
Initial\_Walk\_Speed (fps)
& 20.611& 3.367& 520.241& 0.071\\ \hline
Wait\_Time\_Duration (s)
& 20.016& 3.267& 520.836& 0.074
\\ \hline
Using\_Mobile\_Duration \_At\_Waitpoint (s)
& 14.737& 2.381& 526.113& 0.127\\ \hline
EyeTracking\_People (s)
& 5.229& 0.829& 535.621& 0.365\\ \hline
Gender
& 3.074& 0.486& 537.776& 0.488\\ \hline
Age
& 2.450& 0.387& 538.401& 0.536\\ \hline
Number\_Of\_Glance\_ During\_Crossing
& 0.628& 0.099& 540.223& 0.754\\ \hline
EyeTracking\_Car (s)
& 0.596& 0.094& 540.254& 0.761\\ \hline
EyeTracking\_Road\_Crossing (s)
& 0.589& 0.093& 540.261& 0.761\\ \hline
EyeTracking\_Building (s)
& 0.126& 0.019& 540.725& 0.889\\ \hline
\end{tabular}
}
\label{ANOVA Model}
\end{table*}

\subsubsection{Statistical Tests for Data Robustness}
To ensure dataset robustness, parametric tests (t-tests and ANOVA) and non-parametric tests (Wilcoxon and Kruskal-Wallis) were conducted. Results indicated no statistically significant group differences, confirming model stability and suitability for further analysis (Table \ref{Statistical test}).

\begin{table*}[h!]
\centering
\caption{Summary of parametric and non-parametric test results with p-values for scenario comparisons.}
\renewcommand{\arraystretch}{1} 
\scalebox{0.7}{
\begin{tabular}{|>{\raggedright\arraybackslash}p{6cm}|c|>{\raggedright\arraybackslash}p{4cm}|c|}
\hline
\textbf{Parametric Test} & \textbf{p-value} & \textbf{Non-Parametric Test} & \textbf{p-value} \\ \hline

t-test 1 – Phone vs No phone& 0.706& Wilcoxon 1 & 0.552\\ \hline
t-test 2 – Phone vs Safety Measure& 0.322& Wilcoxon 2 & 0.439\\ \hline
t-Test 3 – No Phone vs Safety Measure& 0.234 & Wilcoxon 3 & 0.221\\ \hline
ANOVA & 0.406& Kruskal-Wallis & 0.470\\ \hline

\end{tabular}
}
\label{Statistical test}
\end{table*}

\subsection{Pedestrian Behavior Analysis Across Distraction and Safety Scenarios}

\subsubsection{Mean Crossing Duration Analysis}
\begin{figure}
\centering
\includegraphics[scale=0.4]{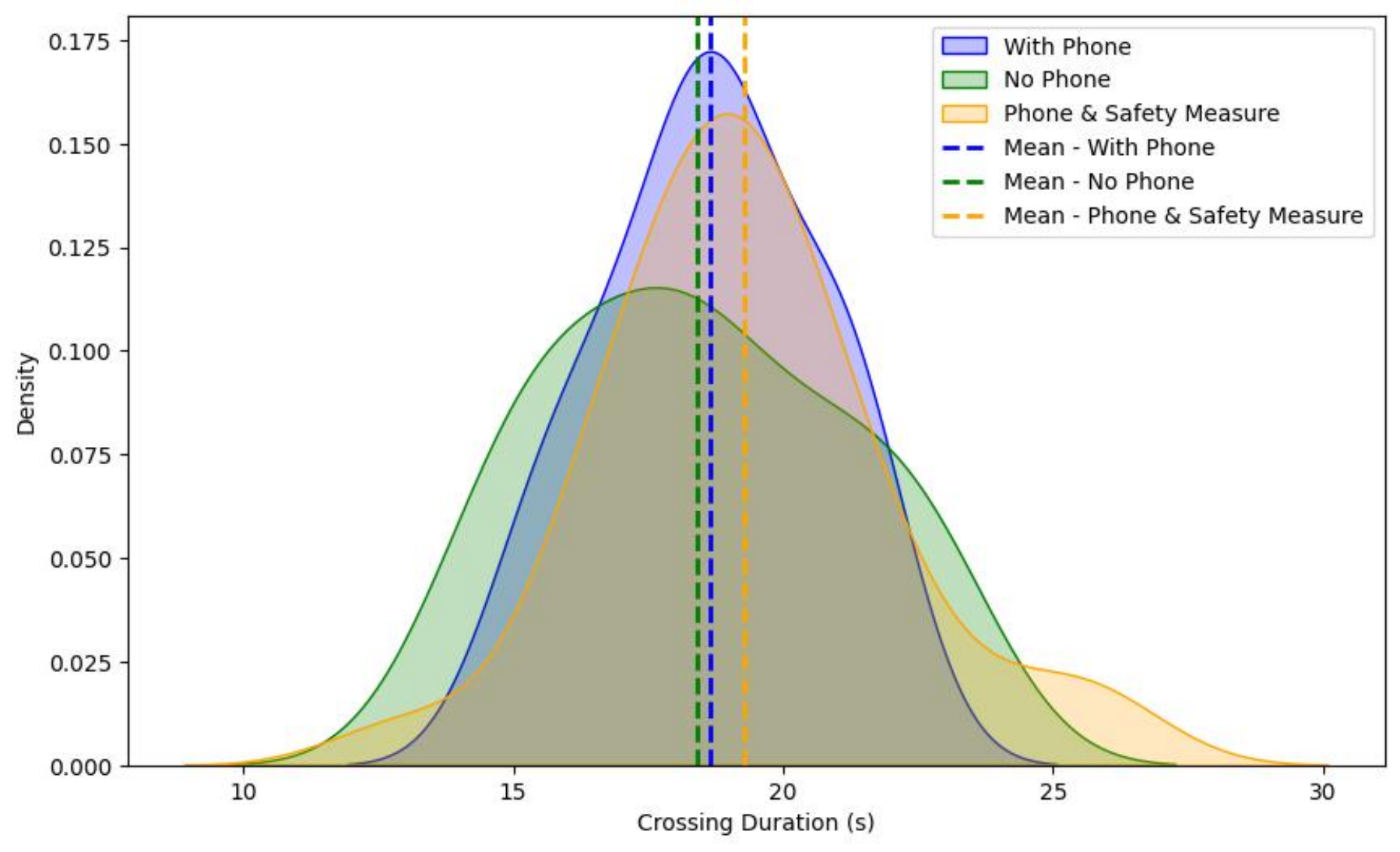}
\caption{Density distributions of pedestrian crossing durations across three scenarios: No Phone, Phone, and Phone \& Safety Measure.}
\label{fig:density-dist}
\end{figure}

Fig. \ref{fig:density-dist} displays the density distributions of crossing durations for three scenarios: No Phone, Phone, and Phone with Safety Measure. The No Phone scenario exhibits the shortest crossing durations, indicating that undistracted pedestrians are the most efficient. In contrast, the Phone scenario demonstrates longer crossing times, reflecting the delays caused by distractions. The Phone \& Safety Measure scenario has the longest durations, suggesting that while LED safety measures provide some benefits, they do not fully mitigate the effects of phone-related distractions.

\subsubsection{Variable Probabilities and Sensitivities in Crossing Duration}
\begin{figure}[H]
\centering
\begin{subfigure}{0.5\columnwidth}
\centering
  \includegraphics[width=0.99\textwidth]{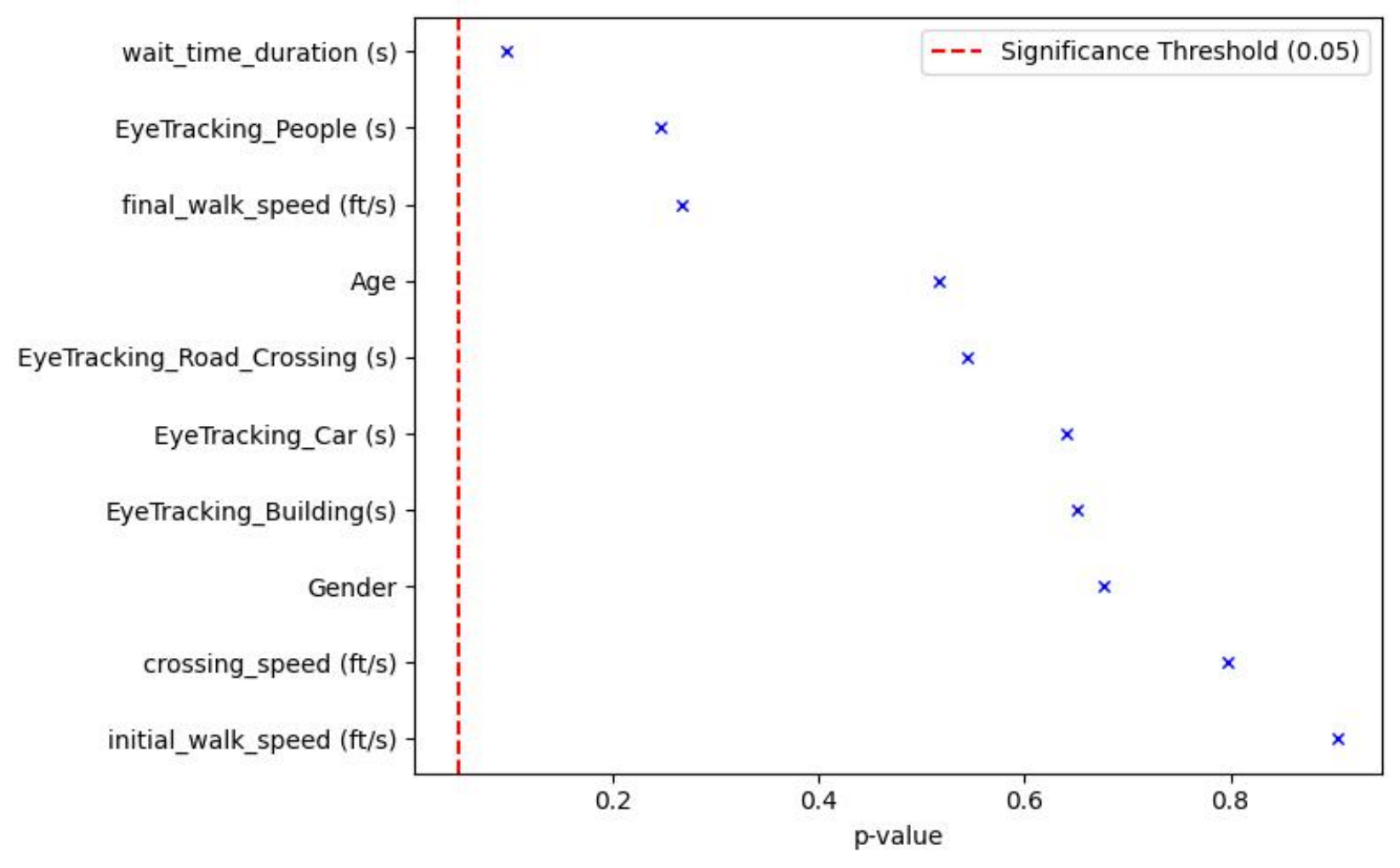}
  \caption{P-value plot for the No Phone scenario, showing variable significance against the 0.05 threshold.}
  \label{fig:no-phone}
\end{subfigure}%
\begin{subfigure}{0.5\columnwidth}
 \centering
\includegraphics[width=0.99\textwidth]{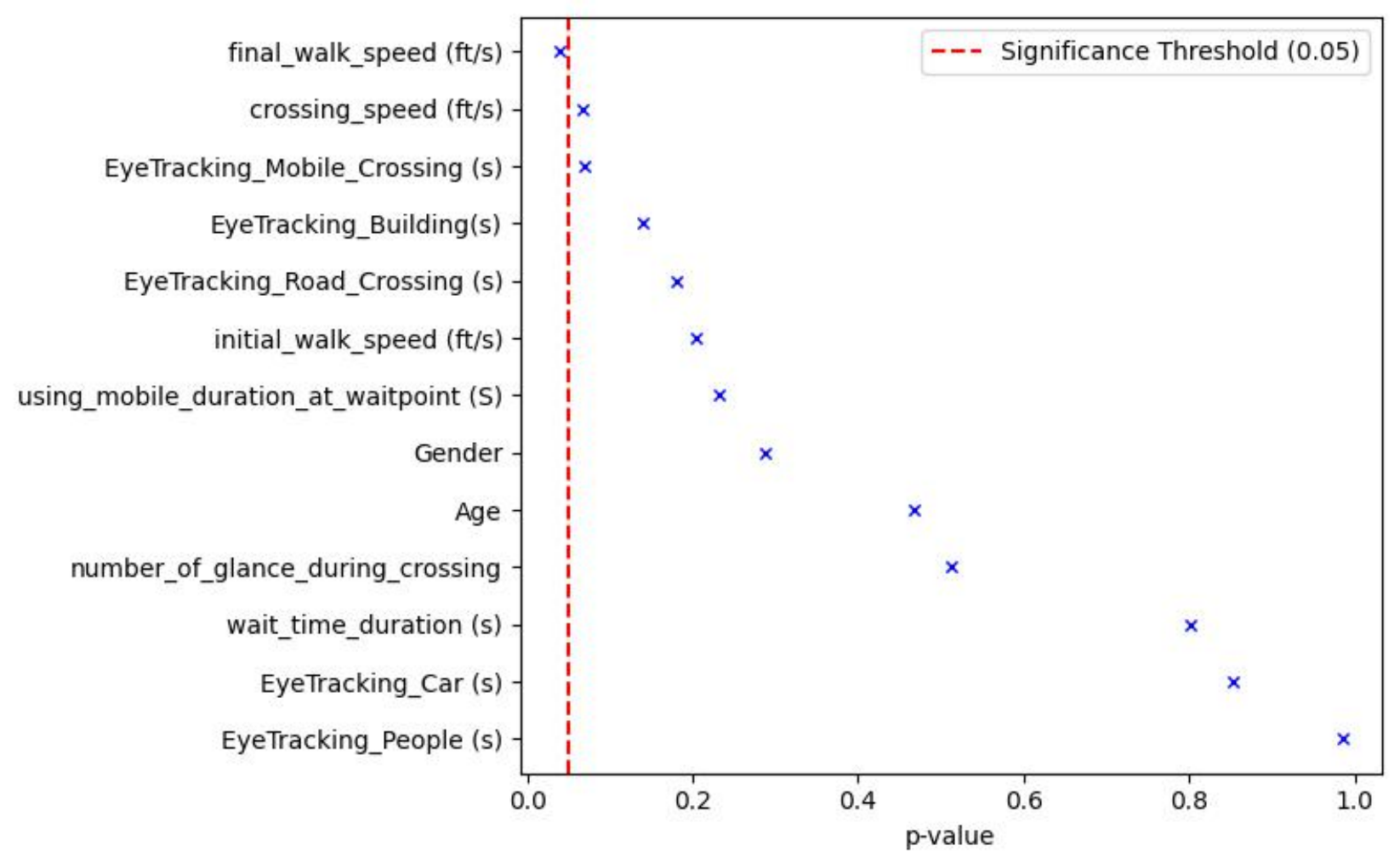}
    \caption{P-value plot for the Phone scenario, showing variable significance against the 0.05 threshold.}
  \label{fig:phone}
\end{subfigure}
\begin{subfigure}{0.5\columnwidth}
 \centering
\includegraphics[width=0.99\textwidth]{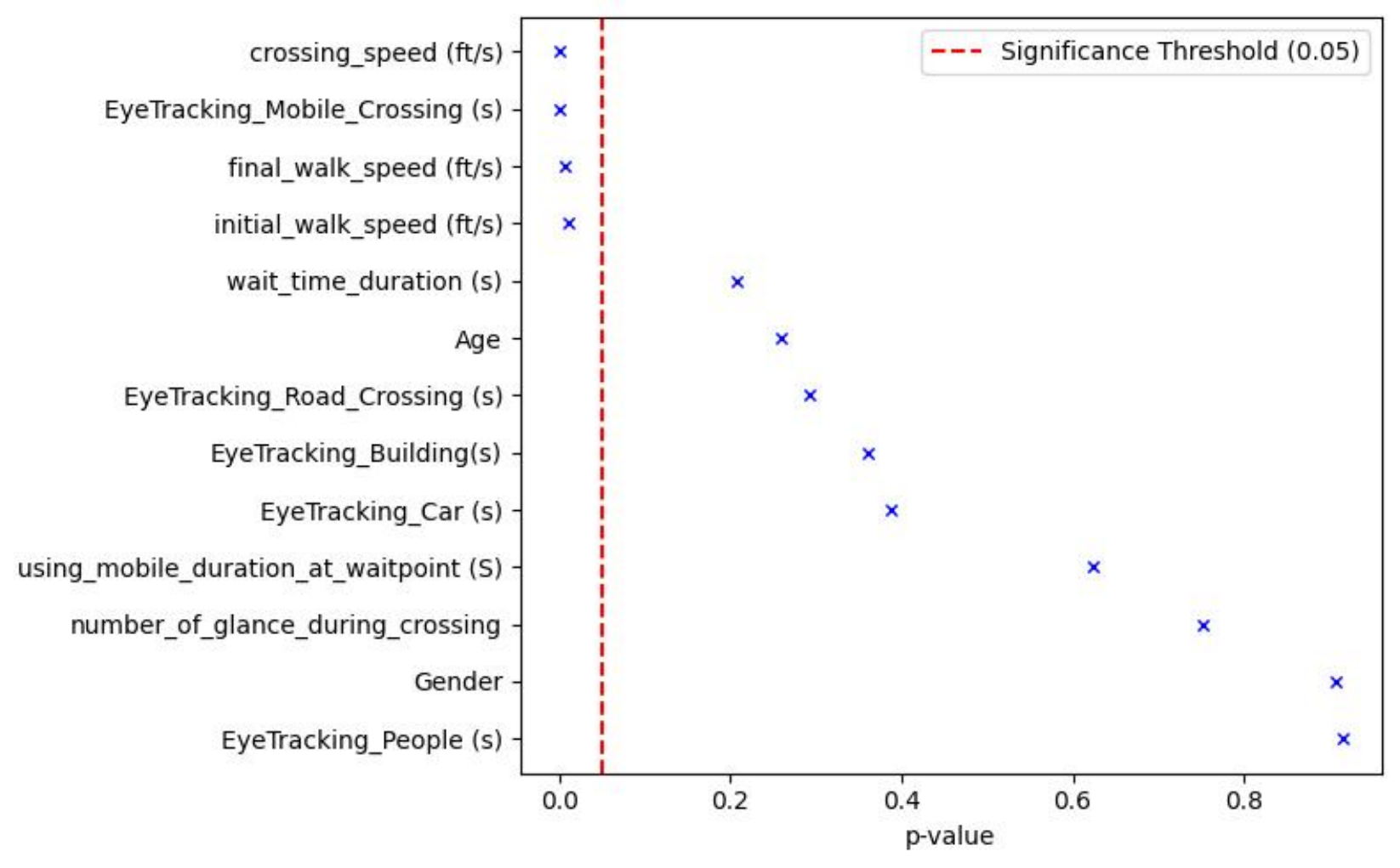}
  \caption{P-value plot for the Phone and Safety measure scenario, showing variable significance against the 0.05 threshold.}
  \label{fig:safety-measure}
\end{subfigure}
\caption{P-values for the Phone and Safety measure scenario}
\end{figure}

Table \ref{Probabilities and Sensitivity} and Figs. \ref{fig:no-phone}, \ref{fig:phone}, and \ref{fig:safety-measure} presents predicted probabilities and sensitivities of key variables on crossing duration. Significant findings include:
\begin{itemize}
    \item No Phone scenario: wait time duration showed a low p-value of 0.09562 and the highest sensitivity at 88\%, indicating that there is a strong influence on pedestrian behavior, though it does not meet the 0.05 significance threshold. Eye tracking on people and final walking speed display notable negative sensitivities (-62\% and -59\%, respectively), suggesting that changes in these variables could inversely affect pedestrian crossing duration. (-14\% and -6\%) for crossing speed and initial walk speed show low sensitivities indicating minimal impact in this scenario. 

    \item Phone Scenario: The analysis of p-values and sensitivity reveals notable impacts of specific variables on pedestrian crossing behavior. Final walk speed variable showed a statistically significant p-value (0.03896) and high negative sensitivity of (-77\%), indicating that as the final walking speed decreases, it strongly impacts crossing duration, likely due to the distraction slowing pedestrians down. Crossing speed is also nearly significant (p-value 0.06693), Eye tracking on Mobile during crossing displays a p-value of (0.06862) and a positive sensitivity of 69\%, suggesting that longer attention to mobile devices while crossing directly affects crossing behavior, likely by prolonging the crossing time. However other variables showed a moderate sensitivity, highlighting some influence on behavior though not statistically significant.  
    \item Phone \& Safety Measure: Crossing speed and Eye Tracking Mobile Crossing have highly significant p-values 0.00009 and 0.00033 and strong sensitivities -177\% and 166\%, correspondingly, which can be explained by the fact that reduced speed and increased attention to mobile devices heavily affects the duration of crossing. Other speed-related factors, like final walk speed and initial walk speed, also contribute noticeably and have a negative sensitivity, which shows that safety measures offset the effects of distraction only moderately.  
\end{itemize}

\begin{table*}[h!]
\centering
\caption{Probabilities and sensitivity scores of variables on pedestrian crossing duration}
\renewcommand{\arraystretch}{1} 
\scalebox{0.7}{
\begin{tabular}{|>{\raggedright\arraybackslash}p{6cm}|c|c|c|c|c|c|}
\hline
 & \multicolumn{2}{|c|}{\textbf{No Phone}} & \multicolumn{2}{|c|}{\textbf{Phone}} & \multicolumn{2}{|c|}{\textbf{Phone \& Safety Measure}} \\ \hline
\textbf{Variable} & \textbf{p-value}& \textbf{Sensitivity} & \textbf{p-value}& \textbf{Sensitivity} & \textbf{p-value}& \textbf{Sensitivity} \\ \hline

Age & 0.517& 35\% & 0.467& -28\% & 0.258& -58\% \\ \hline
EyeTracking\_ Building (s) & 0.651& 24\% & 0.141& -56\% & 0.361& 47\% \\ \hline
EyeTracking\_Car (s) & 0.641& 25\% & 0.852& -7\% & 0.387& -45\% \\ \hline
EyeTracking\_ Mobile\_Crossing (s) & N/A & N/A & 0.069& 69\% & 0.003& 166\% \\ \hline
EyeTracking\_ People (s) & 0.246& -62\% & 0.986& -1\% & 0.916& 5\% \\ \hline
EyeTracking\_ Road\_Crossing (s) & 0.545& 33\% & 0.180& 51\% & 0.293& -54\% \\ \hline
crossing\_speed (ft/s) & 0.797& -14\% & 0.067& -69\% & 0.009& -177\% \\ \hline
final\_walk\_speed (ft/s) & 0.266& -59\% & 0.039& -77\% & 0.006& -132\% \\ \hline
initial\_walk\_speed (ft/s) & 0.904& -6\% & 0.205& -49\% & 0.009& -126\% \\ \hline
number\_of\_glance\_ during\_crossing & N/A & N/A & 0.513& -25\% & 0.752& -16\% \\ \hline
using\_mobile\_durati on\_at\_waitpoint (s) & N/A & N/A & 0.232& 46\% & 0.624& 25\% \\ \hline
wait\_time\_duration (s) & 0.096& 88\% & 0.801& 10\% & 0.208& 64\% \\ \hline
\end{tabular}
}
\label{Probabilities and Sensitivity}
\end{table*}

\subsubsection{Predictions for crossing duration for all scenarios}
Simulation results reveal distinct behavioral patterns across the three scenarios. In the No Phone scenario, pedestrians displayed natural and efficient crossing behaviors. The Phone scenario showed significant delays and greater variability in walking speed, highlighting the adverse effects of distraction. The Phone \& Safety Measure scenario reduced delays slightly, but persistent variability suggests that safety measures alone cannot fully address distraction-related risks.

\begin{figure}[H]
\centering
\begin{subfigure}{0.5\columnwidth}
\centering
  \includegraphics[width=0.99\textwidth]{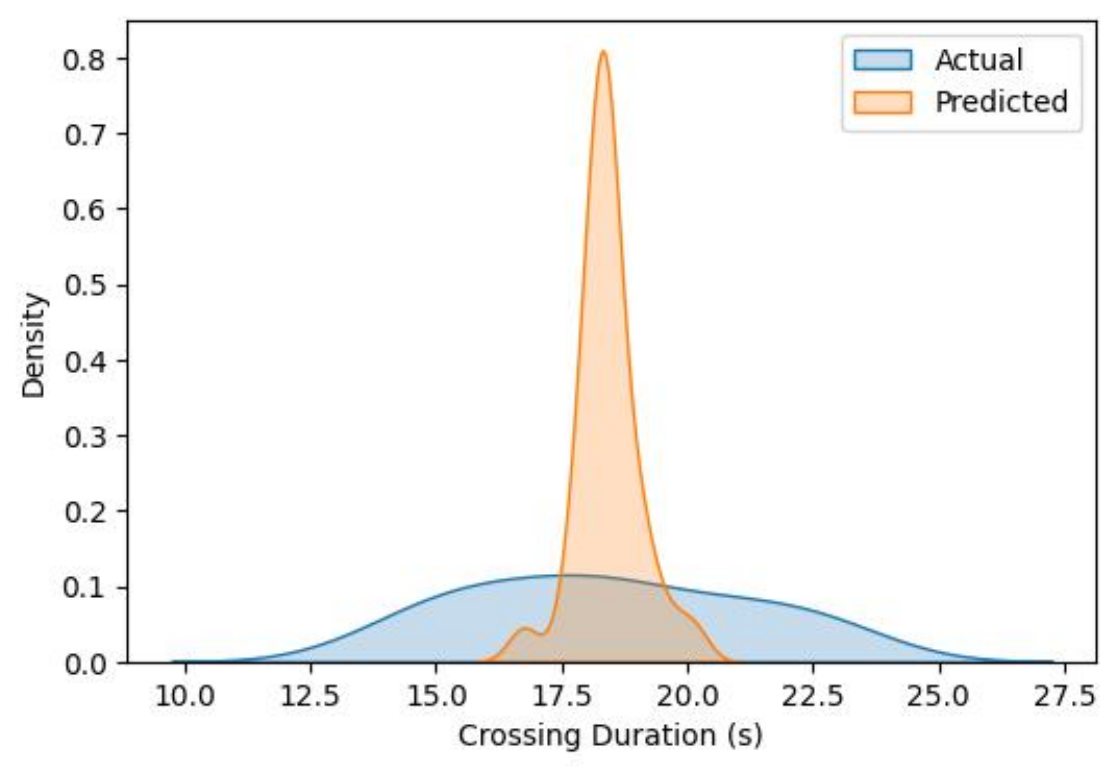}
\caption{Density plot of actual vs. predicted crossing durations for the No Phone scenario.}
  \label{fig:no-phone-scenario}
\end{subfigure}%
\begin{subfigure}{0.5\columnwidth}
 \centering
\includegraphics[width=0.99\textwidth]{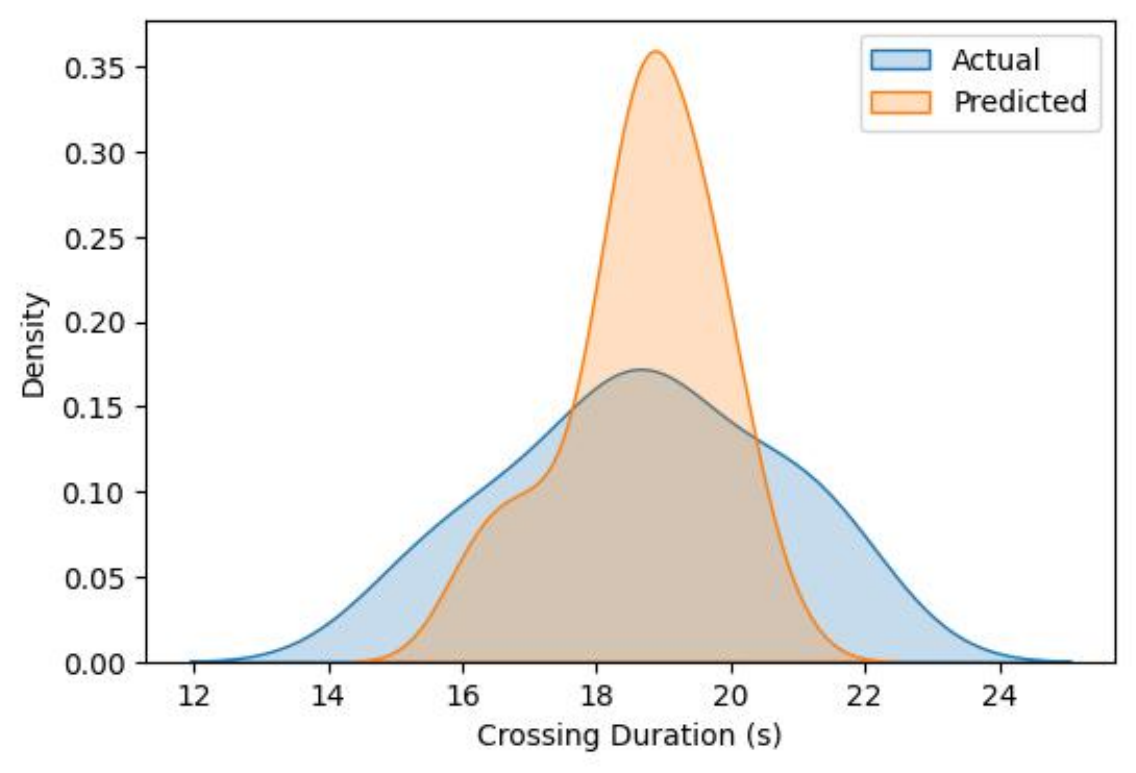}
  \caption{Density plot of actual vs. predicted crossing durations for the Phone scenario.}
  \label{fig:phone-scenario}
\end{subfigure}
\begin{subfigure}{0.5\columnwidth}
 \centering
\includegraphics[width=0.99\textwidth]{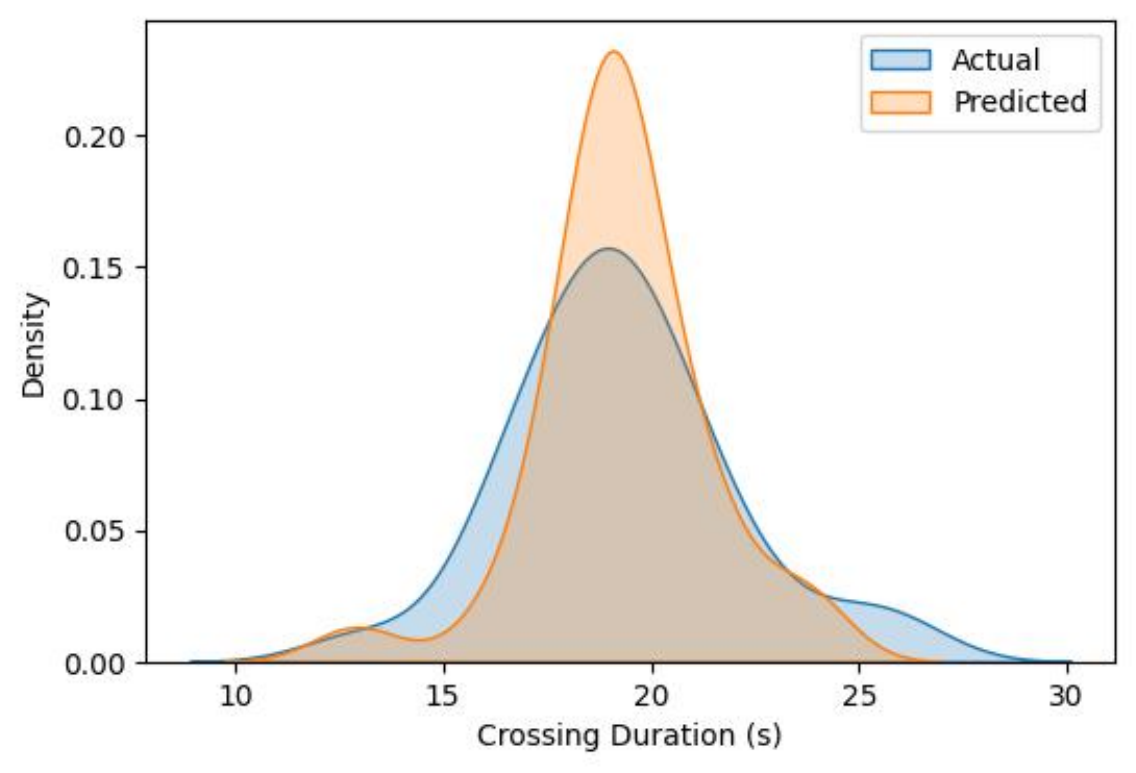}
  \caption{Density plot of actual vs. predicted crossing durations for the Phone and Safety Measure scenario.}
  \label{fig:phone-safety-measure}
\end{subfigure}
\caption{Predicted Distributions}
\end{figure}

In Fig. \ref{fig:no-phone-scenario}, the density plot illustrates the actual and predicted crossing durations for the no-phone scenario. The actual crossing duration shows a broader distribution, spanning from 12.5 to 22.5 seconds, reflecting natural variability in crossing times likely due to differences in walking speeds when pedestrians are undistracted. In contrast, the predicted crossing duration exhibits a narrower peak around 18 seconds. This concentrated range suggests that the model predicts consistent crossing times under no-distraction conditions.

In Fig. \ref{fig:phone-scenario}, the density plot represents crossing durations for pedestrians distracted by phone use. The actual crossing duration ranges from 14 to 24 seconds, with a broader distribution indicating the slower and more variable pace associated with distracted walking. The model predicts crossing durations within a narrower range of 18 to 20 seconds, peaking around 19 seconds. This prediction highlights the moderate delay typically caused by phone use during crossing.

In Fig. \ref{fig:phone-safety-measure}, both actual and predicted crossing durations are shown for the scenario involving phone use with a safety measure. The actual crossing duration ranges widely, from approximately 10 to 30 seconds. Contrary to expectations that the safety measure would reduce crossing durations for distracted pedestrians, the data reveal an extended range of crossing times. This suggests that the safety measure may itself act as a distraction, as supported by findings in Table \ref{Post-survey}. The model predicts a peak crossing duration around 20 seconds, with a relatively narrow distribution compared to the actual data. This peak implies that the model captures a moderate and cautious crossing duration but does not fully account for the variability observed in real-world behaviors when safety measures are introduced.

\subsection{Implications for Urban Planning and Policy}
The findings underscore the need for a combination of technological and behavioral interventions to enhance pedestrian safety, especially at high-risk intersections. While physical safety measures like LED lights have some effectiveness, combining them with public awareness campaigns can better address distraction causes. For example, multi-sensory safety measures (e.g., LED indicators with audio cues) may more effectively alert distracted pedestrians.

\textbf{Practical Recommendations for Urban Planners:}
\begin{itemize}
    \item Implement multi-sensory safety interventions combining visual and auditory signals.
    \item Develop awareness campaigns on the risks of phone use while crossing, collaborating with educational institutions.
    \item Utilize VR-based simulations for testing safety interventions prior to real-world application.
\end{itemize}

\section{Conclusion and Future Work}
This study provides valuable insights into the impact of mobile phone distraction on pedestrian behavior at signalized crosswalks, utilizing an immersive virtual reality (VR) application to evaluate the efficacy of safety interventions. Results demonstrate that mobile phone distraction significantly extends crossing time, reduces situational awareness, and increases variability in crossing speed. While LED-illuminated crosswalks offered some mitigation, they did not fully counteract the adverse effects of distraction. These findings underscore the need for an integrated approach to pedestrian safety that combines physical safety measures with behavioral interventions addressing the root causes of distracted walking. Furthermore, the successful use of VR in simulating realistic, variable conditions highlights its potential as a valuable tool for urban planning and traffic safety research.

To advance this research, our future work would focus on enhancing VR-based interventions by incorporating multi-sensory cues, such as audio and haptic feedback, to more closely simulate real-world environments. Expanding the participant demographic beyond students to include diverse age groups, drivers, and individuals from various regions would increase the generalizability of the findings, capturing a broader range of pedestrian behaviors. Additionally, our future studies would examine a wider array of distractions, such as listening to music, walking in groups, or crossing with pets, as suggested in post-study feedback. These efforts aim to provide a more comprehensive understanding of pedestrian behavior under distraction, ultimately informing more effective safety strategies for diverse urban settings.

\section*{Credit authorship contribution statement}
\noindent \textbf{Methusela Sulle}: Conceptualization, Writing, and Data analysis. \textbf{Judith Mwakalonge}: Writing, data analysis, methodology, and conceptualization. \textbf{Gurcan Comert}:  Writing, data analysis, methodology, and conceptualization. \textbf{Saidi Siuhi}: Writing, data analysis, methodology, and conceptualization. \textbf{Nana Kankam Gyimah}: Writing, methodology, and conceptualization. \textbf{Jaylen Roberts}: Writing and conceptualization. \textbf{Denis Ruganuza}: Writing and data Collection

\section*{Declaration of Competing Interest}
The authors declare that there are no competing financial interests or personal relationships that could have influenced the work reported in this paper.

\section*{Acknowledgements}
This research was supported by the U.S. Department of Education through
Grant No. P382G320015, administered by the Transportation
Program at South Carolina State University (SCSU), the Center for Connected Multimodal Mobility (C2M2) and the National Center for Transportation Cybersecurity and Resiliency (TraCR), USA, headquartered at Clemson University, Clemson, South Carolina, Department of Energy Minority Serving Institutions Partnership Program (MSIPP) managed by the Savannah River National Laboratory under BSRA contract TOA 0000525174 CN1, MSEIP II Cyber Grants: P120A190061, P120A210048, FM-MHP-0678-22-01-00, USA, National Science Foundation (NSF), USA Grants Nos. 1954532, 2131080, 2200457, OIA-2242812, 2234920, and 2305470. Any opinions, findings, conclusions, or recommendations expressed in this material are those of the authors and do not necessarily reflect the views of the C2M2 or TraCR and the official policy or position of the USDOT/OST-R, or any State or other entity. The U.S. Government assumes no liability for the contents or use thereof.

\section*{Data availability}
The data used in this study was collected through experiments designed and conducted by the research team. Due to privacy and ethical considerations involving human participants, the data cannot be shared publicly. However, interested researchers may contact the corresponding author to request access, subject to approval and compliance with relevant data protection regulations.

\bibliography{sample}

\end{document}